\documentclass[fleqn,usenatbib]{mnras}
\usepackage{newtxtext,newtxmath}
\usepackage[T1]{fontenc}

\usepackage{graphicx}	
\usepackage{amsmath}	
\usepackage{ulem}

\title[BH seeds growth with JWST]{Seeking the growth of the first black hole seeds with JWST}

\author[Trinca et al.]{Alessandro Trinca$^{1,2,3}$\thanks{E-mail: alessandro.trinca@inaf.it},
Raffaella Schneider$^{1,2,3,4}$,
Roberto Maiolino$^{5,6,7}$, Rosa Valiante$^{2,3}$,
\newauthor Luca Graziani$^{1,3}$, Marta Volonteri$^{8}$
\\
$^{1}$Dipartimento di Fisica, ``Sapienza'' Universit$\grave{a}$ di Roma, Piazzale Aldo Moro 2, 00185 Roma, Italy \\
$^{2}$INAF/Osservatorio Astronomico di Roma, Via di Frascati 33, 00040 Monte Porzio Catone, Italy \\
$^{3}$INFN, Sezione Roma1, Dipartimento di Fisica, ``Sapienza'' Universit$\grave{a}$ di Roma, Piazzale Aldo Moro 2, 00185, Roma, Italy \\
$^{4}$Sapienza School for Advanced Studies, Viale Regina Elena 291, 00161 Roma, Italy\\
$^{5}$Kavli Institute for Cosmology, University of Cambridge, Madingley Road, Cambridge CB3 0HA, UK \\
$^{6}$Cavendish Laboratory, University of Cambridge, 19 JJ Thomson Avenue, Cambridge CB3 0HE, UK\\
$^{7}$Department of Physics and Astronomy, University College London, Gower Street, London WC1E 6BT, UK\\
$^{8}$Institut d’Astrophysique de Paris, Sorbonne Université, CNRS, UMR 7095, 98 bis bd Arago, 75014 Paris, France
}
\date{Accepted 2022 December 16. Received 2022 December 13; in original form 2022 November 2}

\pubyear{2022}

\begin{document}
\label{firstpage}
\pagerange{\pageref{firstpage}--\pageref{lastpage}}
\maketitle

\begin{abstract}
In this paper we provide predictions for the BH population that would be observable with planned JWST surveys at $5 \le z \le 15$. We base our study on the recently developed Cosmic Archaeology Tool (CAT), which allows us to model BH seeds formation and growth, while being consistent with the general population of AGNs and galaxies observed at $4 \le z \le 7$.
We find that JWST planned surveys will provide a complementary view on active BHs at $z > 5$, with JADES-Medium/-Deep being capable of detecting the numerous BHs that populate the faint-end of the distribution, COSMOS-Web sampling a large enough area to detect the rarest brightest systems, and CEERS/PRIMER bridging the gap between these two regimes. The relatively small field of view of the above surveys preferentially selects BHs with masses $6 \leq {\rm Log} (M_{\rm BH}/M_\odot) < 8$ at $7 \le z < 10$, residing in relatively metal poor (${\rm Log} (Z/Z_\odot) \ge -2$) and massive ($8\leq {\rm Log} (M_*/M_\odot) < 10$) galaxies. At $z \ge 10$, only JADES-Deep will have the sensitivity to detect growing BHs with masses $4 \leq {\rm Log} (M_{\rm BH}/M_\odot) < 6$, hosted by more metal poor ($-3 \leq {\rm Log} (Z/Z_\odot) < -2$) and less massive  ($6 \leq {\rm Log} (M_*/M_\odot) < 8$) galaxies. In our model, the latter population corresponds to heavy BH seeds formed by the direct collapse of super-massive stars in their earliest phases of mass growth. Detecting these systems would provide invaluable insights on the nature and early growth of the first BH seeds.
\end{abstract}




\section{Introduction}

Hundreds of luminous quasars powered by gas accretion onto $> 10^8 \, M_\odot$ supermassive black holes (SMBHs) have been discovered at $z > 6$ (see the recent review by \citealt{inayoshi2020}), with a handful of them at $z > 7$ \citep{mortlock2011, wang2018, matsuoka2019, yang2020}, including the two most distant ones at $z \sim 7.6$ \citep{banados2018, wang2021}. 
Forthcoming surveys have the potential to extend the detection of quasars to fainter luminosities and higher redshifts, 
up to $z \sim 7 - 9$ \citep{wang2017, wang2019, fan2019, barnett2019}.
The origin and evolution of these SMBHs in the first Gyr of cosmic history is deeply connected to the nature of the first low metallicity star forming regions, where the first black hole (BH) seeds are expected to form \citep{omukai2008, volonteri2010, valiante2016, latif2016a, woods2019, volonteri2021, sassano2021}. In addition, the physical processes that enable their early growth are driven by the assembly of the first cosmic structures and are also shaping the properties of their host galaxies \citep{valiante2017, habouzit2020, trinca2022, spinoso2022}. Despite intensive observational efforts, the emission from growing BH seeds has never been directly detected, and the population of low-mass BHs at $z > 6$ has been so far very elusive, possibly because they are too small and faint to be detectable \citep{Volonteri2017, valiante2018observability, valiante2021}, or because of their very low active fraction ($f_{\rm act} \sim 1 \%$), which is the result of short, super-Eddington growth episodes \citep{pezzulli2017a}, or inefficient gas accretion onto low-mass galaxies. In addition, during their efficient growth phase, the progenitors of $z \sim 6$ quasars are expected to be heavily dust-obscured and therefore extremely faint at optical and near-IR wavelengths, transitioning to optically luminous quasars by expelling gas and dust \citep{hopkins2007, Li2008, valiante2011, valiante2014, ginolfi2019}. Detecting these transitioning systems is very challenging, with so far only one convincing candidate detected at $z = 7.2$ \citep{fujimoto2022}.

Constraints on the AGN luminosity function (LF) at $z \geq 6$ have been obtained by the the Subaru High-z Exploration of Low-Luminosity Quasars (SHELLQs) project \citep{matsuoka2018, matsuoka2019} and by deep X-ray surveys \citep{fiore2012, parsa2018, giallongo2019, vito2018}. Extending these data to fainter luminosities is key to constrain the nature of the first BH seeds and dominant growth mode, as suggested by a number of independent studies \citep{ricarte2018a, ricarte2018b, piana2021, trinca2022}. In particular, \citet{trinca2022} suggest that current data appear to favour models where the growth of BH seeds is Eddington-limited or where super-Eddington accretion occurs via a slim disc during gas-rich galaxy mergers. At $z \leq 6 - 7$ the main difference between these two models is strongly related to the efficiency of growth of light BH seeds, formed as remnants of the first stars \citep{valiante2016, trinca2022}. In the reference model, these light BH seeds fail to grow and a clear gap appears at the low-mass and low-luminosity end of the mass and luminosity functions, which are dominated by BHs descending from heavy seeds that originate by the direct collapse of super-massive stars. 
Conversely, in the second model light and heavy BH seeds are able to grow in gas-rich galaxy mergers, resulting into a BH mass and luminosity distributions where BH descendants of both light and heavy seeds can contribute to the same mass and luminosity bins.

While detecting systems in the luminosity range where these differences appear will be challenging even with JWST, at $z > 6$ the observation of a slower decline in the number density of bright AGNs might be a hint that early BH evolution is strongly driven by short period of enhanced accretion occurring during galaxy mergers \citep{trinca2022}, or favored by
the presence of dense gas clumps \citep{bournaud2011, lupi2014} and by low angular momentum gas inflows \citep{dubois2012}.

A number of studies have shown that JWST would be able to detect the first growing BH seeds \citep{pacucci2015, natarajan2017}, the most luminous of which up to $z \sim 16$ \citep{barrow2018, valiante2018observability} and beyond \citep{whalen2020}. Discriminating the nature of these growing BH seeds is more challenging, as light BH seeds and heavy BH seeds power very similar emission spectra when they are luminous enough to exceed JWST sensitivity limits \citep{valiante2018observability}. In addition, searches for growing BH seeds must be extended out to $z > 10$, when the probability of observing
them evolving in isolation, before they lose memory of their original nature via galaxy mergers, is expected to be higher \citep{valiante2018statistics}. Even for these luminous, isolated growing BH seeds, it is challenging to disentangle their rest-frame UV/optical emission from that of the stellar component, and thus properly designed colour-colour selections will be required \citep{natarajan2017unveiling, Volonteri2017, valiante2018observability, Nakajima2022, Goulding2022, inayoshi2022}.

Motivated by these results, in this paper we predict, based on our semi-analytical model CAT \citep{trinca2022}, the number of accreting BHs that would be observable with planned surveys with JWST.  In particular, we adopt the CAT reference model assuming that nuclear BHs grow without exceeding the Eddington limit from a population of light and heavy BH seeds formed in the first metal-poor galaxies ($z \geq 15$),
and where only heavy seeds are able to grow efficiently in mass powering luminous AGNs at $4 < z \leq 6 - 7$.

The paper is organized as follows: in Section \ref{sec:Model} we briefly summarize the main features of the model. In Section \ref{sec:AGNobs} we discuss the observability of accreting BHs at different redshifts by planned JWST surveys, while Sections \ref{sec:BHmass} and \ref{sec:BHmetal} discuss the typical BH masses, luminosities and metallicities of their host galaxies. Finally, in Section \ref{sec:conclusions} we summarize and discuss our main conclusions.

\section{The Cosmic Archaeology Tool}
\label{sec:Model}

In this work we characterize the evolution of high-redshift galaxies and black holes using the Cosmic Archaeology Tool (CAT), originally presented in \citet{trinca2022}. Here we briefly summarize the main features of the model, and we refer interested readers to the original paper for a thorough description of the model and the adopted calibration of free parameters.

CAT is a semi-analyical model which has been developed to describe the formation of the first galaxies and BHs and follows their co-evolution through cosmic times. We rely on the galaxy formation model \textsc{galform} \citep{parkinson2008}, which is based on the Extended Press Schechter formalism, to reconstruct a large sample of dark matter hierarchical merger histories representative of the evolution of the entire galaxy population between $z=4$ and $z=24$. We adopt a mass resolution that corresponds to a virial temperature of $T_{\rm vir} = 1200 {\rm K}$, so that we can describe star formation occurring in molecular and atomic-cooling halos, corresponding to virial temperatures $1200 {\rm K} \leq T_{\rm vir} < 10^4 {\rm K}$ and $T_{\rm vir} \geq 10^4 {\rm K}$, respectively. The minimum resolved halo mass therefore ranges from $9.4 \times 10^5 \, \rm M_\odot$ at $z \sim 24$ to $1.0 \times 10^7 \, \rm M_\odot$ at $z \sim 4$.
Following halo virialization, the gas gets accreted, 
cools, and triggers star formation. Inside each galaxy, the star formation rate (SFR) is computed as:
\begin{equation}
{\rm SFR} = f_{\rm cool} \, M_{\rm gas} \, \epsilon_{\rm SF} / \tau_{\rm dyn},
\label{eq:SFR}
\end{equation}
where $M_{\rm gas}$ is the available gas mass, $\epsilon_{\rm SF}$ is the star formation efficiency and $\tau_{\rm dyn} = [R_{\rm vir}^3 / (G \, M_{\rm halo}) ]^{1/2}$ is the halo dynamical time. The SF efficiency $\epsilon_{\rm SF} = 0.05$ represents a free parameter of the model. The factor $f_{\rm cool}$ quantifies the reduced cooling efficiency due to Lyman Werner (LW)  radiation (corresponding to photon energies $11.2 - 13.6 \, \rm{eV}$), which can photodissociate molecular hydrogen, suppressing gas cooling in molecular-cooling halos. Following \citet{valiante2016, debennassuti2017, sassano2021}, the value of $f_{\rm cool}$ depends on the halo virial temperature, redshift, gas metallicity and intensity of the illuminating LW radiation. Conversely, in atomic cooling halos we set $f_{\rm cool} = 1$. 

We assume that the first (Pop III) stars are characterized by a top-heavy Initial Mass Function (IMF), that we parametrize as a Larson IMF:
\begin{equation}
\Phi (m_*) \propto m_*^{\alpha -1} \, e^{-m_*/m_{\rm ch}}
\label{eq:IMF}
\end{equation}
where $\alpha = -1.35$, $m_{\rm ch} = 20\, M_\odot$ and the possible range of stellar masses is $10 \, M_\odot \leq m_* \leq 300\, M_\odot$ \citep{debennassuti2014, debennassuti2017}. During each SF episode, 
we stochastically sample the IMF and we compute the emitted stellar radiation, supernova (SN) explosion rate, $R_{\rm SN} (t)$, metal and dust yields and final BH masses in a self-consistent way \citep{trinca2022}.

When the gas metallicity of star forming regions exceeds a critical value of $\rm{Z_{cr}} = 10^{-3.8} \, \rm{Z_\odot}$, metal-fine structure lines and dust cooling increase the cooling efficiency \citep{omukai2001, schneider2002, omukai2005, schneider2006, schneider2012b}, leading to a transition in the characteristic stellar masses. We therefore assume that above $\rm{Z_{cr}}$, Pop II stars form in the mass range $0.1 \, M_\odot \leq m_* \leq 100\, M_\odot$ according to a Larson IMF with $m_{\rm ch} = 0.35\, M_\odot$ \citep{debennassuti2014, debennassuti2017}. We then follow their emitted stellar radiation, metal and dust yields evolving each stellar population on its characteristic evolutionary timescales (i.e. we do not assume an instantaneous recycling approximation). For a thorough description of the two-phase interstellar medium (ISM) and chemical evolution module implemented in CAT we refer to \citet{valiante2014, debennassuti2014}.

Black hole evolution is described starting from a seeding prescription which is consistent with the above baryonic evolution. Following each Pop III star 
formation episode, we assume that the heaviest among the newly formed BH remnants settles 
in the center of the galaxy, forming a light BH seed. Heavy BH seeds with a mass of $10^5 \, \rm{M_\odot}$ form by the so-called Direct Collapse (DC) mechanism, when the gas collapses almost iso-thermally with no fragmentation, leading to the formation of a single super massive star that becomes unstable, due to nuclear exhaustion or GR instabilities \citep{hosokawa2012, inayoshi2014, latif2013, ferrara2014, becerra2015, latif2016a, becerra2018}. This formation pathway operates inside atomic-cooling halos (where $T_{\rm vir} \geq 10^4 \, \rm{K}$), when metal and dust cooling is still inefficient ($\rm{Z} \leq \rm{Z_{cr}}$) and when molecular cooling is suppressed by a strong illuminating LW flux. The latter condition is usually expressed as $\rm{J_{LW}} \geq \rm{J_{cr}}$, where $\rm{J_{LW}}$ is the cumulative flux into the LW energy band in units of $10^{-21} \, \rm{ erg \, s^{-1} \, cm^{-2} \, Hz^{-1} \,sr^{-1}}$. The value of $\rm{J_{cr}}$ is still very uncertain (see the recent reviews by \citealt{woods2019} and \citealt{inayoshi2020} for a thorough discussion and
\citealt{chon2020} for more recent results). 
Following \citet{trinca2022}, here we adopt a threshold value of $\rm{J_{cr}} = 300$. Also, we do not consider the possibility to form intermediate mass BH seeds from runaway mergers in dense stellar clusters (see  \citealt{sassano2021} for a recent investigation that considers all the three BH seeds populations).

Note that our results rely on the assumptions that BH seeds, once formed, settle at the center of the host galaxy. 
High resolution zoom-in simulations show that if the BH seed mass is less than $10^5 \rm \, M_\odot$, its dynamical evolution is very perturbed by the irregular stellar distribution in high-redshift galaxies \citep{pfister2019}. This effect will further suppress the growth of light BH seeds, but should have a smaller impact on the observable population of accreting BHs that we have analysed, which descend from heavy BH seeds.

Once formed, seed BHs can grow through gas accretion and mergers with other BHs. The gas accretion rate onto BHs is described by the Bondi-Hoyle-Lyttleton (BHL) formula \citep{hoyle1941, bondi1952}:
\begin{equation}
    \dot{M}_{\rm BHL} = \alpha \frac{4 \pi G^2 M^2_{\rm BH}\rho_{\rm gas}(r_{\rm A})}{c^3_{\rm s}},
\end{equation}
\noindent
where $c_{\rm s}$ is the sound speed and $\rho_{\rm gas}(r_{\rm A})$ is the gas
density evaluated at the radius of gravitational influence of the BH, $r_{\rm A} = 2 G M_{\rm BH}/c_{\rm s}^2$. 
The $\alpha$ parameter, which is not present in the original Bondi formula, is introduced to account for the enhanced gas density in the inner regions around the central BH, and it is one of the free parameters of the model. Following the model calibration presented in \citet{trinca2022}, a value of $\alpha = 90$ is shown to reproduce the observed population of high redshift quasars at $z \gtrsim 6$.

In our reference model we assume that the gas accretion rate, $\dot{M}_{\rm accr}$, cannot exceed the Eddington limit, so that:
\begin{equation}
\dot{M}_{\rm accr} = {\rm min} (\dot{M}_{\rm BHL}, \dot{M}_{\rm Edd}),
\end{equation}
\noindent
and the BH mass growth rate is computed as:
\begin{equation}
\dot{M}_{\rm BH} = (1 - \epsilon_{\rm r}) \dot{M}_{\rm accr}. 
\end{equation}
\noindent
In the above expressions, $\dot{M}_{\rm Edd} = L_{\rm Edd}/(\epsilon_{\rm r} c^2)$, $\epsilon_{\rm r} = 0.1$ is the adopted radiative efficiency, and $L_{\rm Edd} = 4 \pi c G M_{\rm BH} m_{\rm p}/\sigma_{\rm T}$ is the Eddington luminosity ($c$ is the speed of light, $m_{\rm p}$ is the proton mass and $\sigma_{\rm T}$ is the Thomson scattering cross section).

Following \citet{valiante2011} we assume that two BHs coalesce during major mergers, i.e. if the mass ratio of their interacting host DM halos is $\mu > 1/10$ \citep{tanaka2009}. In our model, both the host galaxies and their nuclear BHs merge within the characteristic time interval of the simulation ($\Delta t \sim 0.5 - 4 \, \rm Myrs$) and the merger product settles in the nuclear region of the final galaxy. Conversely, in minor mergers ($\mu < 1/10$), only the most massive among the two nuclear BHs is assumed to migrate in the center of the newly formed galaxy. 
As shown by \citet{tremmel2018}, the sinking timescale of nuclear BHs after a galaxy merger might be significantly longer than the one assumed in CAT, although mergers are expected to be more efficient if BH seeds have masses $\sim 10^5 M_\odot$ and they are hosted in galaxies with high central stellar and gas densities, which facilitate BH binary formation and hardening \citep{volonteri2020}.
A future version of CAT will implement time delays between the merger of the host galaxies and the subsequent coupling of their nuclear BHs, potentially leading to a merger. While this approach is of crucial importance to investigate the rate of nuclear BH merger events expected through cosmic times \citep{valiante2021}, it should have a smaller impact on the predictions presented in this work. The growing, and thus detectable, BH population is dominated by gas accretion on heavy seeds which have stable dynamics; furthermore in our model BH-BH mergers have a small contribution to growth as also suggested by previous works \citep{dubois2014,valiante2016,pacucci2020}.

The abundance of gas inside each galaxy is affected by photo-heating feedback, which - at each given redshift - suppresses star formation in haloes with virial temperatures below the temperature of the intergalactic medium (IGM, \citealt{valiante2016})\footnote{We consider $T_{\rm IGM} = Q_{\rm HII} \, T_{\rm reio} + (1 - Q_{\rm HII}) \, T_{\rm HI}$, where $T_{\rm reio} = 2 \times 10^4 \rm \, K$, $T_{\rm HI} = 0.017 (1+z)^2$ and the filling factor of HII regions, $Q_{\rm HII}$, is computed as in \citep{valiante2016}.}, and by mechanical feedback due to galaxy-scale outflows driven by the energy released by supernova (SN) explosions and BH accretion, 
\begin{equation}
\dot{M}_{\rm{ej}} = \dot{M}_{\rm{ej, SN}} + \dot{M}_{\rm{ej, AGN}} 
\label{eq:fbk}
\end{equation}
where $\dot{M}_{\rm ej,SN}$ and $\dot{M}_{\rm ej, AGN}$ are the SN- and AGN-driven outflow rates. The first term is defined as:
\begin{equation}
\dot{M}_{\rm{ej, SN}} = \frac{2 E_{\rm SN} \epsilon_{\rm w,SN} R_{\rm SN} (t)}{v_{\rm e}^2} ,
\label{eq:SNfbk}
\end{equation}
where $E_{\rm SN}$ represents the explosion energy per SN and $R_{\rm SN} (t)$ is the SN explosion rate, which depend on the SF history and on the nature of the stellar populations hosted by each galaxy\footnote{For Pop III stars,  $E_{\rm SN}$ is assumed to be $2.7\times 10^{52}$\, erg, while for Pop II/I stars, $E_{\rm SN} = 1.2\times 10^{51}$ erg.}, $v_{\rm e} = (2 G M/R_{\rm vir} )^{1/2}$ is the escape velocity of the galaxy, and $\epsilon_{\rm w,SN} = 1.6 \times 10^{-3}$ is a free parameter representing the SN-driven wind efficiency.

The second term in Eq. \ref{eq:fbk} is computed as,
\begin{equation}
\dot{M}_{\rm{ej, AGN}} = 2 \, \epsilon_{\rm{w, AGN}} \, \epsilon_r \, \dot{M}_{\rm accr} \, \biggl( \frac{c}{v_{\rm e}} \biggl)^2.
\label{eq:AGNfbk}
\end{equation}
\noindent
where $\epsilon_{\rm w,AGN}$ is the AGN-driven wind efficiency. Following \citet{trinca2022}, in our reference model, we assume that $\epsilon_{\rm w,AGN} = 2.5 \times 10^{-3}$.

The bolometric luminosity of each accreting BH, $L_{\rm bol} = \epsilon_{\rm r} \dot{M}_{\rm accr} c^2$ can then be converted into a B-band ($4400 $\AA) luminosity using the bolometric correction proposed by \citet{duras2020}, $L_{\rm bol}/L_{\rm B} = 5.13$, and extrapolated to other optical/UV wavelengths assuming a power law slope with $L_\nu \propto \nu^{-0.44}$ \citep{trinca2022}. 
Finally, following \citet{merloni2014}, we assume the fraction of obscured AGNs to be a decreasing function of their intrinsic X-ray luminosity,
\begin{equation}
    f_{\rm obs} = 0.56 + \frac{1}{\pi} {\rm arctg} \, \biggl( \frac{43.89 - {\rm Log} \,L_{\rm X}} {0.46} \biggl)
\end{equation}
\noindent
where $L_{\rm X}$ is expressed in erg/s and it is computed from the bolometric luminosity adopting the bolometric correction proposed by \citet{duras2020}. This model has been shown to provide a good description of the currently observed galaxy and AGN populations at $4 \le z \leq 7$ and to comply with global constraints on the cosmic star formation rate density evolution and on the history of cosmic reionization (see \citealt{trinca2022} and Trinca et al. in prep).


\begin{table}
    \centering
    \caption{Properties of planned JWST surveys. We report the name of the survey, the area coverage, the AB limiting magnitude at 10$\sigma$ in the F200W filter. Note: $^*$ COSMOS-WEB does not use the F200W filter, so the limiting magnitude shown in the table is an interpolation between the F150W and the F277W limiting magnitudes.}
\begin{tabular}{l|l|l}
Survey & Area [arcmin$^2$] & limiting mag \\
\hline
JADES-Deep     &  46 & 29.9\\
JADES-Medium     & 190 & 29.0\\
CEERS & 97 & 28.2 \\
PRIMER  & 695 & 27.7\\
COSMOS-WEB & 2180 & 27.1$^*$ \\
\end{tabular}
\label{table:surveys} 
\end{table}


\begin{figure*}
	\includegraphics[width=\linewidth]{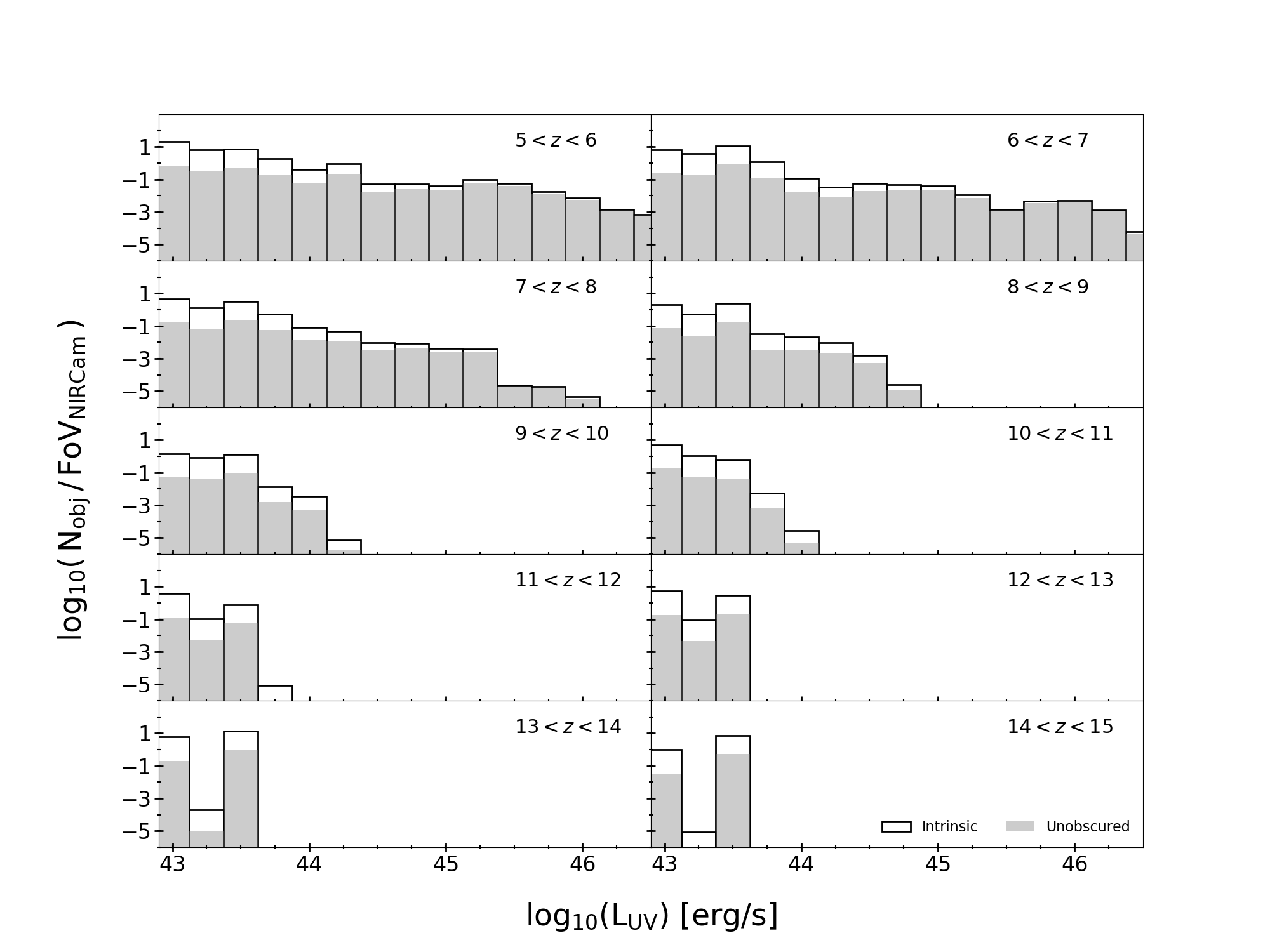}
    \caption{Number of BHs per NIRCam field of view as a function of their UV luminosity for different redshift bins in the range $5 < z <15$. Empty histograms show the total (obscured and unobscured) population, while filled grey histograms represent the unobscured systems.}
    \label{fig:Ncount_Luv_tot}
\end{figure*}


\begin{figure*}
	\includegraphics[width=\linewidth]{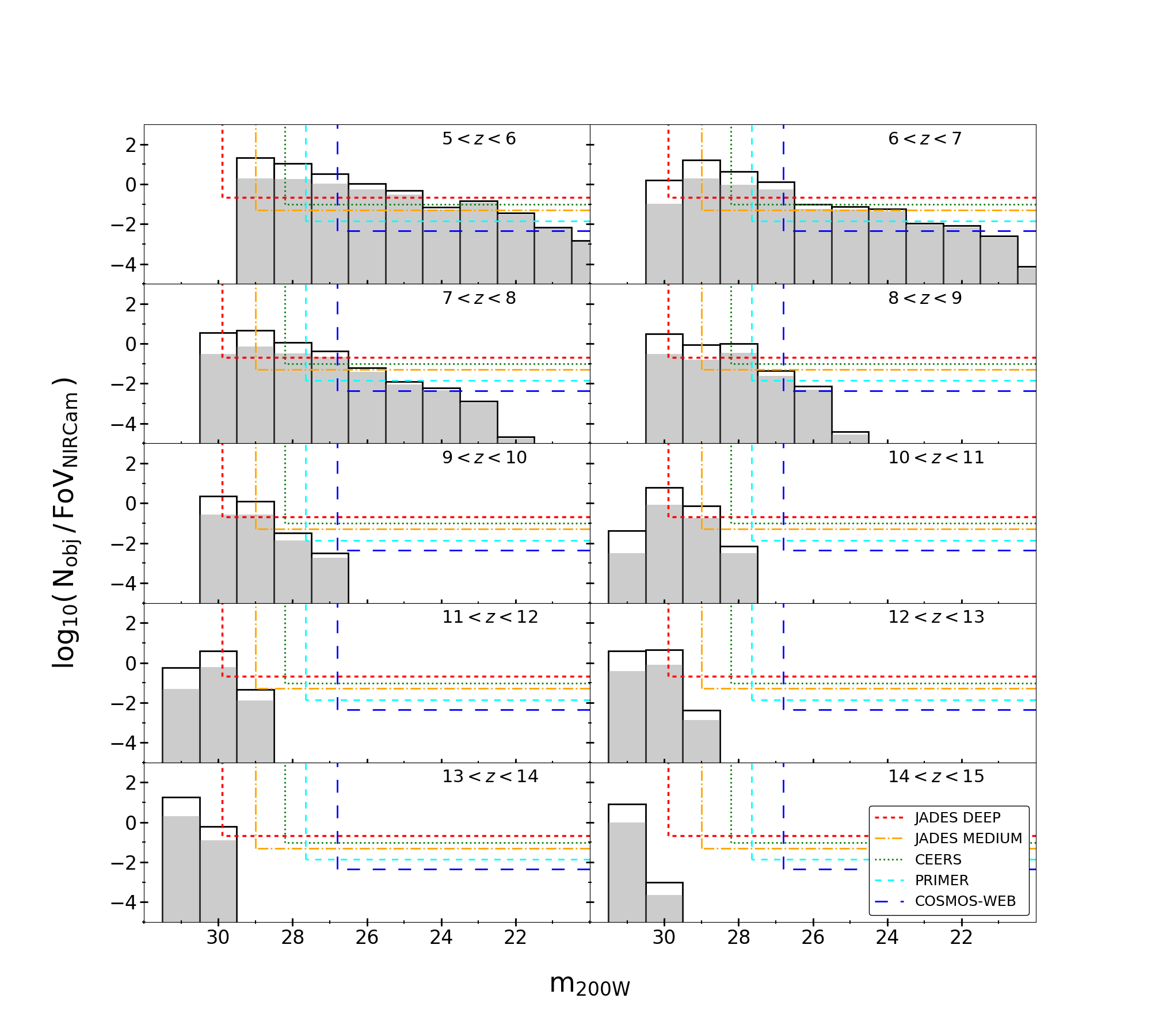}
    \caption{Same as Fig. \ref{fig:Ncount_Luv_tot} but represented as a function of the AB magnitude in JWST F200W NIRCam filter. Vertical colored lines represent the limiting magnitude for different JWST surveys. Horizontal lines show instead the density limit to observe at least one object in the area covered by the survey.}
    \label{fig:Ncount_Mag_tot}
\end{figure*}


\section{RESULTS}

In this section, we illustrate our main findings. We first present the predicted number of accreting BHs that would be observable in planned JWST surveys at different redshifts. Then, we discuss the physical properties of the observable BH population and of their hosts. In particular, we focus on the BH mass and on the stellar mass and metallicity of the host galaxy. By combining these properties we hope to identify BHs in their early phases of mass growth, i.e. close to their formation epochs, and hence to constrain the nature of the first BH seeds.

\subsection{The observability of accreting BHs}
\label{sec:AGNobs}
In Figs. \ref{fig:Ncount_Luv_tot} and \ref{fig:Ncount_Mag_tot} we show the number of accreting BHs expected per NIRCam field of view as a function of their UV luminosity and F200W magnitude as predicted by the CAT reference model. The distributions are shown for different redshift bins in the range $5 < z < 15$. Vertical dashed lines in Fig. \ref{fig:Ncount_Mag_tot} indicate the sensitivity limits of planned surveys with JWST; horizontal dashed lines with the same color show instead the number density of BHs that corresponds to  having at least 1 system in the volume of the survey. On this regard, Table \ref{table:surveys} provides a summary of the assumed area and 10$\sigma$ limiting magnitude of each survey. For the limiting magnitude we are considering 10$\sigma$ as simple detection is generally not enough, and reliable colors, and possibly spectra, are needed to discriminate candidate AGNs from other populations of galaxies.

We find that planned surveys should be able to detect growing BHs up to $z \sim 12 - 13$. At $z \leq 7 - 8$, a complementary view of the accreting BH populations will be provided by different surveys, with JADES-Medium/-Deep being capable of detecting the numerous BHs that populate the faint-end of the distribution, COSMOS-Web sampling a large enough area to detect the rarest brightest systems, and CEERS/PRIMER bridging the gap between these two regimes. At higher redshift, it will become progressively more challenging for shallower surveys to detect accreting BHs, and at $z > 10 - 11$ only JADES-Deep will be sensitive enough to potentially detect growing BHs. 

The expected properties of accreting BHs with luminosity above the sensitivity limit of each survey are reported in Table \ref{table:results}, where we split the population in three redshift bins: $5 \le z < 7$, $7 \le z < 10$, and $z \geq 10$. For each survey, we provide information on the expected number of detectable BHs with and without obscuration correction, their UV luminosity, the range of BH masses, and the metallicity and stellar mass of their host galaxy. These latter properties are discussed in more details in the following subsections.


\begin{figure*}
	\includegraphics[width=\linewidth]{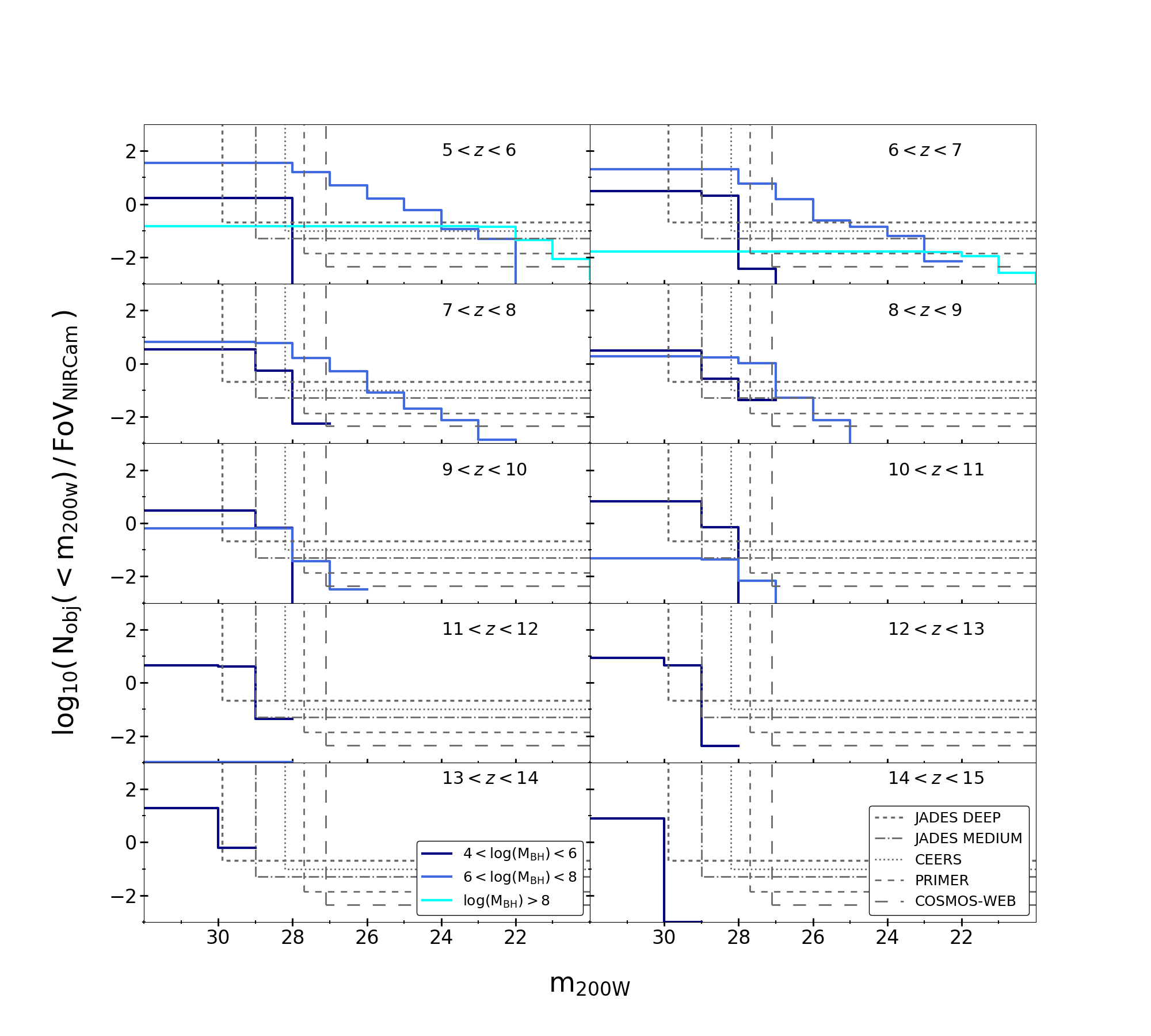}
    \caption{Number of accreting BHs brighter than $m_{\rm 200W}$ per NIRcam field of view as a function of their apparent magnitude $m_{\rm 200W}$ assuming negligible obscuration for different redshift bins in the range $5 < z < 15$. Here we split the BH population in three separate bins of BH mass, represented by different colored lines (see the legenda). As in Fig. \ref{fig:Ncount_Mag_tot} vertical and horizontal lines represent the magnitude and density limit for different surveys: JADES-Deep (short-dashed), JADES-Medium (dot-dashed), CEERS (dotted), PRIMER (dashed), COSMOS-WEB (long dashed). The BH population that would be observable by each survey depends on its area and sensitivity, with COSMOS-Web sampling a large enough area to detect a few of the rarest systems with ${\rm Log} (M_{\rm BH}/M_\odot) > 8$ at $z < 7$, and JADES-Deep being capable of detecting the heavy BH seeds with masses $4 < {\rm Log} (M_{\rm BH}/M_\odot) < 6$ out to $z \sim 13-14$.}
    \label{fig:Ncount_mass}
\end{figure*}


\begin{figure*}
	\includegraphics[width=\linewidth]{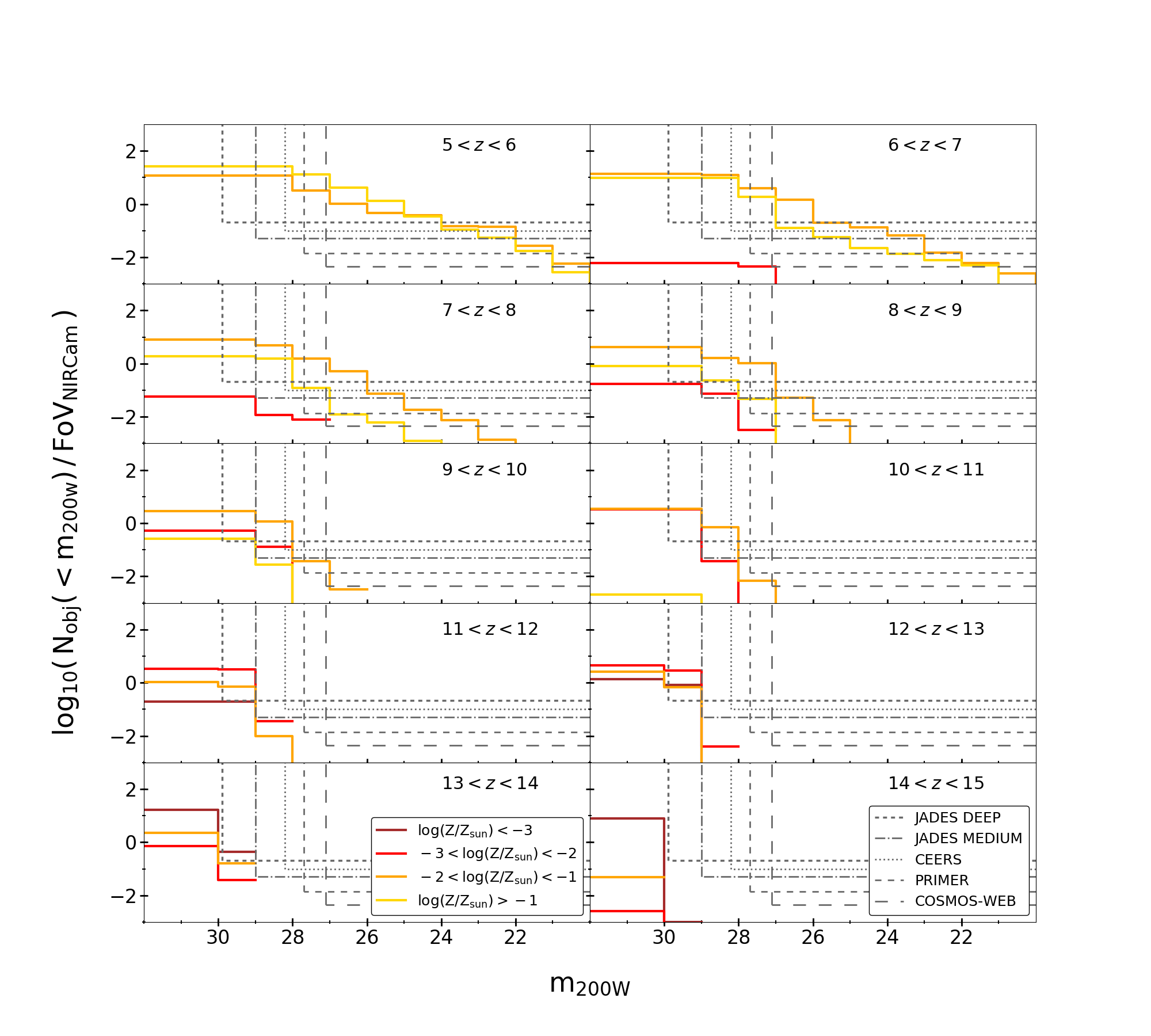}
    \caption{Same as Fig. \ref{fig:Ncount_mass}, but with colored lines representing different metallicity bins for the BH host galaxy. At $z < 10$ observable BHs are preferentially hosted by galaxies with ${\rm Log} (Z/Z_\odot) \ge -2$, while at higher redshifts an increasing fraction of observable systems resides in galaxies with $-3 \leq {\rm Log} (Z/Z_\odot) < -2$.}
    \label{fig:Ncount_metal}
\end{figure*}


\begin{figure*}
	\includegraphics[width=\linewidth]{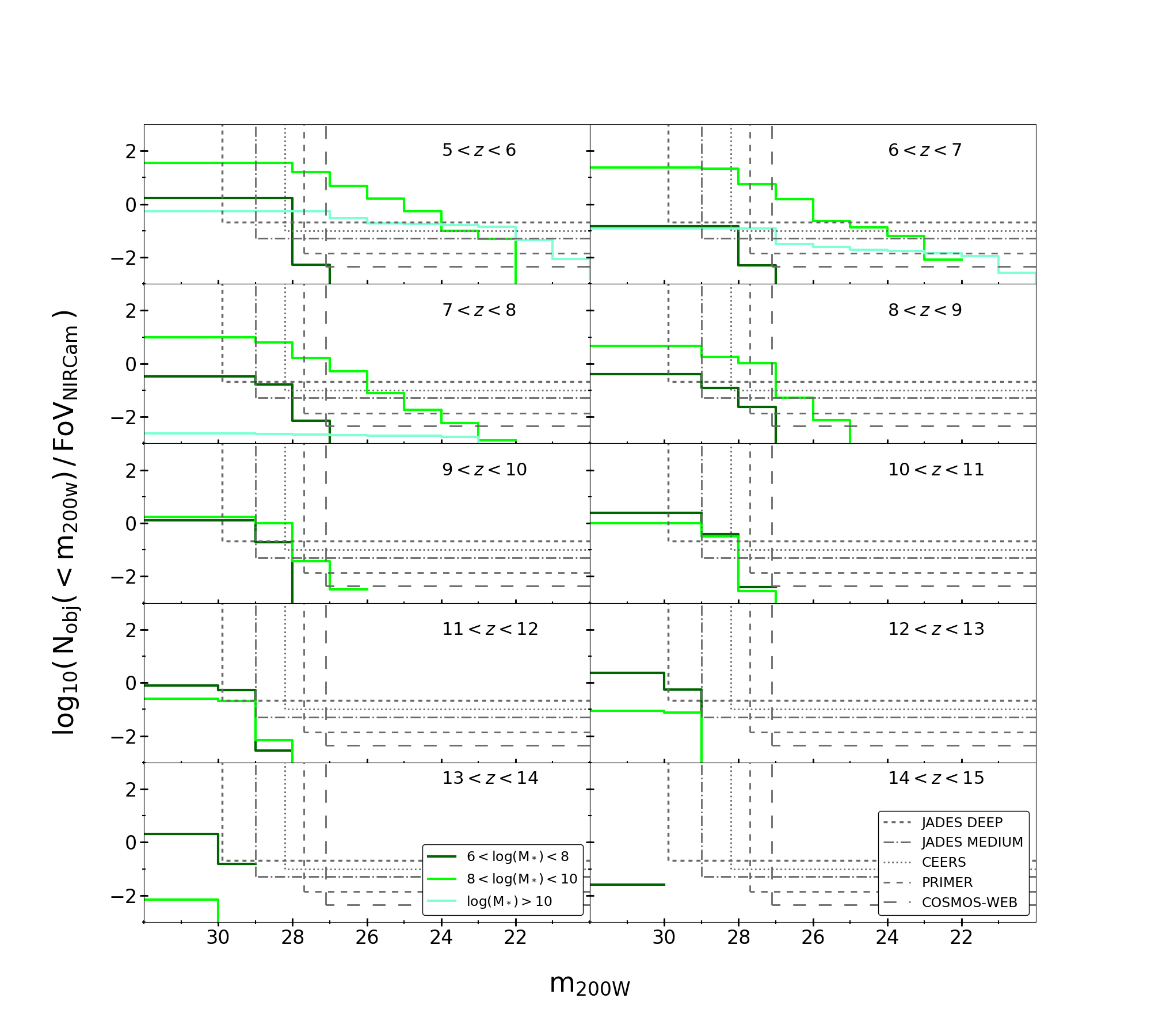}
    \caption{Same as Fig. \ref{fig:Ncount_mass}, but with colored lines representing different bins of stellar mass of the BH host galaxy. At redshift $z < 10$, the large majority of the observable BH population is hosted by galaxies with stellar mass $ 8 \leq {\rm Log} (M_*/M_\odot) < 10$. At earlier times an increasing fraction of systems resides in smaller galaxies with $6 \leq {\rm Log} (M_*/M_\odot) < 8$, despite their observability is only within the reach of deeper surveys as JADES-Deep and JADES-Medium.}
    \label{fig:Ncount_StMass}
\end{figure*}


\begin{figure*}
	\includegraphics[width=0.9\linewidth]{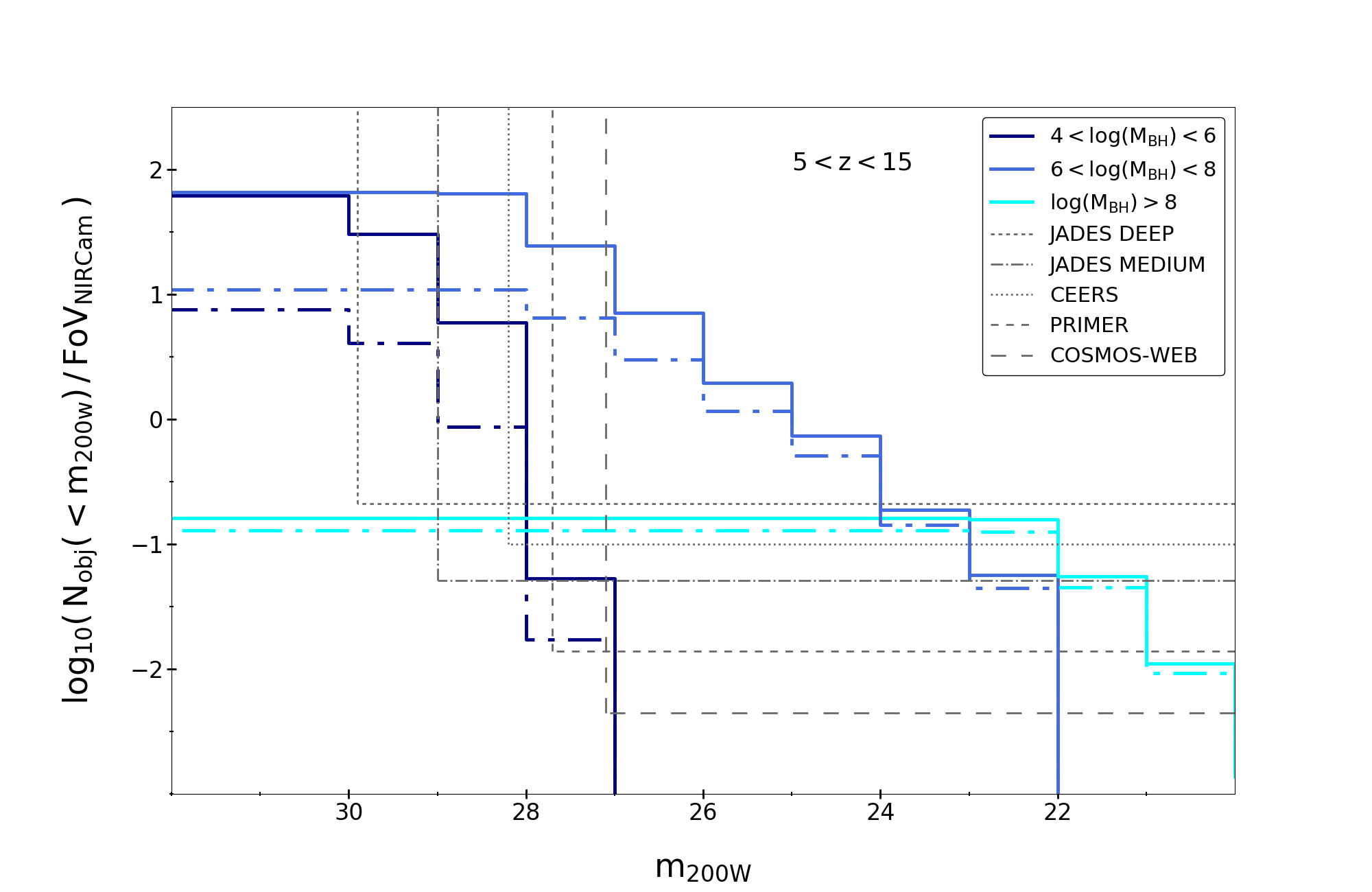}
    \caption{Expected number of accreting BHs brighter than $m_{\rm 200W}$ per NIRcam field of view as a function of their apparent magnitude $m_{\rm 200W}$ in the redshift range $5 < z < 15$. The population is split in bins of BH mass, represented by different colored lines, with thick solid (loosely dash-dotted) lines neglecting (adopting) an obscuration correction. As in Fig. \ref{fig:Ncount_Mag_tot} vertical and horizontal lines represent the magnitude and density limit for different surveys: JADES-Deep (short-dashed), JADES-Medium (dot-dashed), CEERS (dotted), PRIMER (dashed), COSMOS-WEB (long dashed).}
    \label{fig:Ncount_mass_tot}
\end{figure*}


\begin{figure*}
	\includegraphics[width=0.9\linewidth]{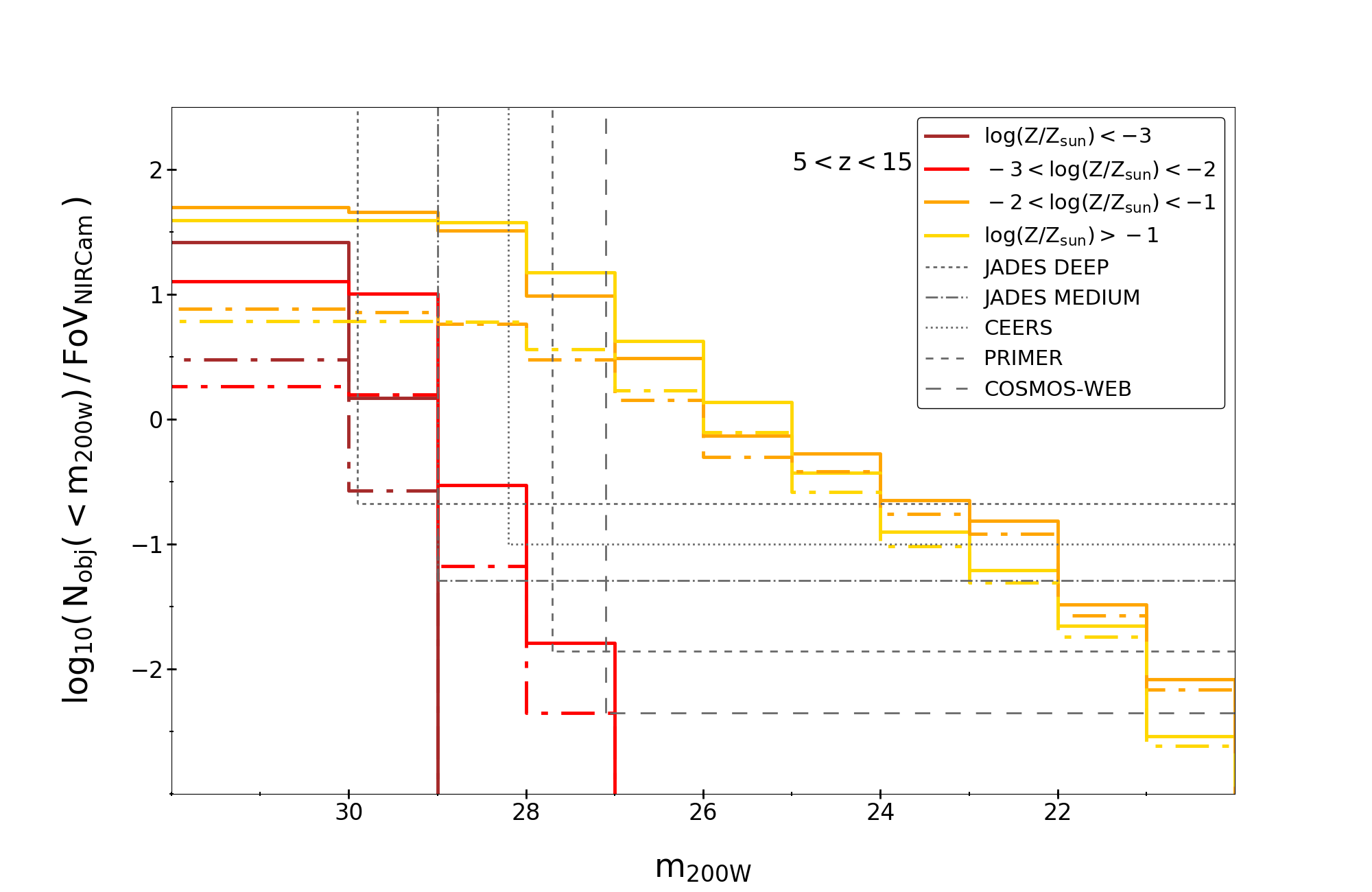}
	\includegraphics[width=0.9\linewidth]{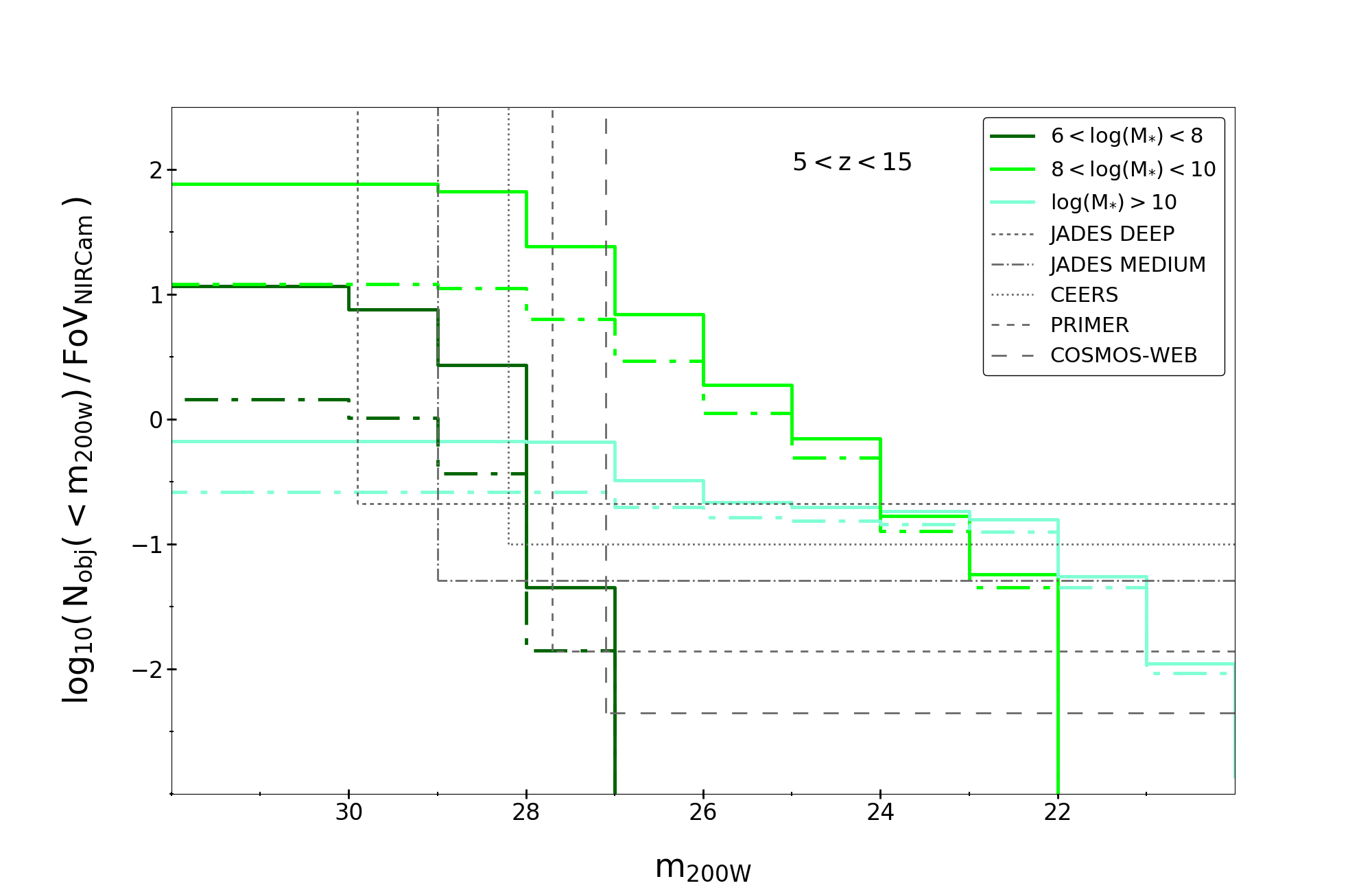}
    \caption{Same as Figure \ref{fig:Ncount_mass_tot} but the BH population is split in bins of host galaxy metallicity (upper panel) and stellar mass (lower panel).The thick solid (loosely dash-dotted) lines show the expected number of systems neglecting (adopting) an obscuration correction.}
    \label{fig:Ncount_Mstar_tot}
\end{figure*}


\begin{table*}
    \centering
    \caption{Properties of accreting BH populations observable by different JWST surveys in the redshift ranges $5 \leq z  \leq 7$, $7  \leq z  \leq 10$ and $z \geq 10$. For each survey, we report the expected number of observable systems (in parenthesis we give the number when an obscuration correction is applied), the corresponding range of BH mass, BH UV luminosity, host galaxy metallicity and stellar mass.}
\begin{tabular}{l|c|c|c|c|c}
Survey & $\rm N_{BH} \, (unobscured)$ &  $\rm M_{BH} \, [M_\odot]$ & $\rm Log \, L_{UV}/[erg/s]$ & $\rm Z \, [Z_\odot]$ & $\rm M_* \, [M_\odot]$ \\
\hline
& & $  \qquad\qquad\qquad\qquad\qquad \rm 5 \leq z \leq 7$ & &\\
\hline
JADES-Deep     &  287 (45) & $[10^4-10^8]$ & $[43.0 - 44.9]$ & $> 10^{-2}$  & $>10^6$\\
JADES-Medium   &  804 (149)& $[10^4-10^8]$ & $[43.3 - 45.7]$ & $> 10^{-2}$  & $>10^6$\\
CEERS          & 175 (48)  & $[10^6-10^8]$ & $[43.7 - 45.7]$ & $> 10^{-2}$  & $>10^8$\\
PRIMER         & 701 (243) & $>10^6$       & $[44.1 - 46.1]$ & $> 10^{-2}$  & $>10^8$\\
COSMOS-WEB     & 1086 (491)& $>10^6$       & $[44.3 - 46.5]$ & $> 10^{-2}$  & $>10^8$\\
\hline
& & $ \qquad\qquad\qquad\qquad\qquad \rm 7 \leq z \leq 10$ & &\\
\hline
JADES-Deep    &  63 (12) & $[10^4-10^8]$ & $[43.2 - 44.3]$ & $> 10^{-3}$ & $[10^6-10^{10}]$\\
JADES-Medium  & 122 (30) & $[10^4-10^8]$ & $[43.5 - 44.8]$ & $> 10^{-3}$ & $[10^6-10^{10}]$\\
CEERS         & 21 (8)   & $[10^6-10^8]$ & $[43.9 - 44.4]$ & $> 10^{-2}$ & $[10^8-10^{10}]$\\
PRIMER        & 73 (31)  & $[10^4-10^8]$ & $[44.4 - 44.8]$ & $> 10^{-2}$ & $[10^8-10^{10}]$\\
COSMOS-WEB    & 86 (45)  & $[10^6-10^8]$ & $[44.5 - 45.6]$ & $> 10^{-2}$ & $[10^8-10^{10}]$\\
\hline
& & $  \qquad\qquad\qquad\qquad\qquad \rm z \geq 10$ & &\\
\hline
JADES-Deep    &  32 (5) & $[10^4-10^6]$ & $[43.4 - 43.6]$ & $[<10^{-3}-10^{-1}]$ & $[10^6-10^{10}]$ \\
JADES-Medium  & 8 (2)   & $[10^4-10^6]$ & $[43.8 - 43.9]$ &  $[10^{-2}-10^{-1}]$ & $[10^6-10^{10}]$ \\
CEERS         & <1 (<1) & / & / & / & / \\
PRIMER        & <1 (<1) & / & / & / & / \\
COSMOS-WEB    & <1 (<1) & / & / & / & / \\
\hline
\end{tabular}
\label{table:results} 
\end{table*}


\subsection{Masses of the observable BH population}
\label{sec:BHmass}
In Fig. \ref{fig:Ncount_mass} we show the distribution in mass of the accreting BH population as a function of redshift. Each panel represents the cumulative number of objects brighter than a given magnitude in the F200W filter band assuming negligible obscuration (the same population represented by the empty histograms in Figs. \ref{fig:Ncount_Luv_tot} and \ref{fig:Ncount_Mag_tot}), for three different bins of M$_{\rm BH}$: a small mass bin ($4 \leq {\rm Log} M_{\rm BH}/M_\odot < 6$), which sample BH seeds and their early growth phase, an intermediate mass bin ($6 \leq {\rm Log} M_{\rm BH}/M_\odot < 8$), and a large mass bin (${\rm Log} M_{\rm BH}/M_\odot \ge 8$), which sample the upper end of the mass function, where SMBHs powering quasars are expected to reside.
The relatively small field of view of these surveys preferentially selects the most common BH population in each redshift bin, i.e. BHs with masses $6 \leq {\rm Log} (M_{\rm BH}/M_\odot) < 8$ at $5 \le z < 10$, and BHs with masses $4 \leq {\rm Log} (M_{\rm BH}/M_\odot) < 6$ at $z \ge 10$. 

In our reference model, all the observable BH population at $z \ge 10$ correspond to heavy BH seeds formed by the direct collapse of super-massive stars in their earliest phases of mass growth \citep{trinca2022}, due to the inefficient mass growth of light BH seeds, whose maximum mass and UV luminosity are $ \rm M_{BH}\sim 10^{3.5} \, M_\odot$ and $\rm L_{UV} \sim 10^{41} \, erg \, s^{-1}$ at all redshifts. The observable population of heavy BH seeds at $z > 10$ and of their descendants at $z < 10$ accrete gas with rates that typically range between [0.1 - 1] $\dot{M}_{\rm Edd}$. 

According to the statistical analysis presented in \citet{valiante2018statistics}, 
at $z \geq 10$ even BHs that are the direct progenitors of SMBHs powering quasars at $z \sim 6 - 7$, are more likely to be observed evolving in isolation, i.e. before they and their host galaxies experience mergers with other systems. Hence, even in the presumably overdense regions which facilitate the formation of the first quasars and their host galaxies, accreting BHs at $z \ge 10$ are more likely to keep memory of their birth conditions. Detecting these systems would provide invaluable insights on the nature and early growth of the first BH seeds. 

\subsection{Host metallicity of the observable BH population}
\label{sec:BHmetal}
Cosmological models of light, medium weight, and heavy BH seed 
formation show that their birth environments generally require
suppression of efficient gas cooling and fragmentation \citep{omukai2008}. These conditions generally imply metal-poor or metal-free environments, possibly exposed to a strong UV background, which suppress H$_2$ cooling \citep{valiante2016, trinca2022} and favour high gas accretion rates \citep{chon2020, sassano2021}. 
Measuring the metallicity of the host galaxies of high redshift accreting BHs with JWST may provide additional important constraints on BH seeding models. 

In Fig.\ref{fig:Ncount_metal} we show the distribution in metallicity of galaxies hosting the accreting BH population in different redshift bins. Similarly to Fig. \ref{fig:Ncount_mass}, each panel shows the cumulative number of objects brighter than a given magnitude in the F200W filter band assuming negligible obscuration for four different metallicity bins: an extremely metal-poor bin (${\rm Log} Z/Z_\odot < -3$), which sample the formation sites of BH seeds, and three additional bins where $Z/Z_\odot$ is increased by 1 dex ($-3 \leq {\rm Log} Z/Z_\odot < -2$, $-2 \leq {\rm Log} Z/Z_\odot < -1$, ${\rm Log} Z/Z_\odot \ge -1$). The figure shows that at $z < 10$ observable BHs are expected to be preferentially hosted by galaxies with metallicities ${\rm Log} (Z/Z_\odot) \ge -2$, while at higher redshifts an increasing fraction of observable systems are predicted to be hosted by galaxies with $-3 \leq {\rm Log} (Z/Z_\odot) < -2$. 

It is important to stress that, although seeds are expected to form in pristine or extremely metal-poor environments, with $Z  \lesssim 10^{-4} Z_\odot$, once the star formation starts in their host galaxies, metal enrichment proceeds on the short evolutionary timescales of massive stars (a few Myrs). Hence, it is not surprising that even relatively unevolved BH seeds at $z > 10$ are predicted to be 
hosted by galaxies with a broad range of metallicities, as shown in Fig. \ref{fig:Ncount_metal}.
Yet, our results suggest that at $z > 10$ JADES-Deep will be sensitive enough to detect heavy BH seeds in very metal-poor galaxies.

\subsection{Host galaxy stellar mass}
\label{sec:StellarMass}
Characterizing the properties of galaxies hosting the first accreting BH seeds is important to guide the search and detection of these elusive objects.
In Figure \ref{fig:Ncount_StMass} we show how the population of accreting BHs is distributed as a function of the host galaxy stellar mass for different redshift bins in the range $5 < z < 15$. Similarly to Fig. \ref{fig:Ncount_mass}, each panel shows the cumulative number of objects brighter then a given magnitude in the F200W filter band, assuming negligible obscuration, for three different bins of the host galaxy stellar mass: a low-mass bin $ 6 \leq {\rm Log} (\rm M_*/M_\odot) < 8$, sampling unevolved or recently formed galactic systems, and two additional mass bins, $8 \leq {\rm Log} (\rm M_*/M_\odot) < 10$ and ${\rm Log} (\rm M_*/M_\odot) \geq 10$, representing more evolved and massive hosts.
At $z < 10$, our model predicts that observable BHs are preferentially hosted by galaxies with stellar mass $8 \leq {\rm Log} (\rm M_*/M_\odot) < 10$ over a wide range of BH UV magnitudes. Only the brightest systems with magnitude $\rm m_{200W} < 23$ at $z < 7$ are more likely to be hosted by more massive galaxies, with $ {\rm Log} (\rm M_*/M_\odot) > 10$. At $z > 10$, an increasing fraction of accreting BHs is hosted by less evolved systems with $6 \leq {\rm Log} (\rm M_*/M_\odot) < 8$. 
The figure shows that BH seeds hosted in these low-mass galaxies can be potentially detected by JADES-Deep up to $z \sim 12$.

It is important to note that accreting BHs hosted by less massive systems have a larger probability to keep memory of their formation environments, as their evolution has not been significantly perturbed by dynamical interactions with other galaxies \citep[see e.g.][]{valiante2018statistics}.
Our predictions hence suggest that deep JWST surveys have the potential to provide important indications on the formation environment of the first heavy BH seeds.
 
\subsection{The effect of obscuration}
\label{sec:obscuration}
In Fig. \ref{fig:Ncount_mass_tot} we show the number of accreting BHs brighter than a given magnitude $m_{200W}$ expected per NIRCam field of view in the entire redshift range $5<z<15$. The accreting BH populations is split in bins of BH mass, using the same ranges of values adopted in Fig. \ref{fig:Ncount_mass}.
In Fig. \ref{fig:Ncount_Mstar_tot} we show the same distribution but divided in bins of host galaxy metallicity (upper panel) and stellar mass (lower panel). In both Fig. \ref{fig:Ncount_mass_tot} and \ref{fig:Ncount_Mstar_tot} the thick solid (thick dashed) lines show the expected number of systems when no obscuration is assumed (with an obscuration correction). 
It is clear that even when accounting for obscuration JADES-Deep/-Medium can potentially detect a mixed population of accreting BHs, with masses in the small and intermediate mass bins, while surveys like CEERS, PRIMER and COSMOS-Web will target BHs with masses in the intermediate mass bin, except for a small number of bright, massive systems.
Because of the relatively rapid chemical evolutionary timescales, the dominant population of accreting BHs that are potentially detectable with JWST resides in galaxies with metallicities ${\rm Log} (Z/Z_\odot) \ge -2$. Only JADES-Deep may be sensitive enough to detect more metal-poor environments, with a fraction of the faintest accreting BHs residing in galaxies with $-3 \le {\rm Log} (Z/Z_\odot) < -2$, where we should expect dust obscuration to be negligible. Yet, our obscuration correction, which is based on an empirical model calibrated at $0.3 \le z \le 3$ \citep{merloni2014}, does not foresee a metallicity dependence and likely overestimates the fraction of obscured BHs at very low metallicities. We will come back to this point in the next section.
In a similar way, when accounting for obscuration, our predictions suggest that the dominant population of accreting BHs detectable with JWST is hosted by galaxies with stellar masses $8 \leq {\rm Log} (\rm M_*/M_\odot) < 10$, since BH growth is suppressed in lower mass systems, as also suggested by other studies \citep{dubois2015, bower2017, habouzit2017, angles-alcazar2017}. Only deep surveys as JADES-Deep/-Medium may be able to detect fainter BHs residing inside less massive galaxies.

\section{Discussion and conclusions}
\label{sec:conclusions}

In this paper, we have discussed the capabilities of JWST to constrain the population of accreting BHs at $z > 5$. We find that planned JWST surveys may provide a very complementary view on this early BH population, with JADES-Medium/-Deep being capable of detecting the numerous BHs at the faint-end of the distribution, COSMOS-Web sampling a large enough area to detect the rarest brightest systems, and CEERS/PRIMER bridging the gap between these two regimes. At higher redshift, it will become progressively more challenging for shallower surveys to detect accreting BHs, and at $z > 10 - 11$ only JADES-Deep will be sensitive enough to potentially detect growing BHs. 

We find that the relatively small field of view of these surveys preferentially selects the most common BH population in each redshift bin, i.e. BHs with masses $6 \leq {\rm Log} (M_{\rm BH}/M_\odot) < 8$ at $7 \le z < 10$, and BHs with masses $4 \leq {\rm Log} (M_{\rm BH}/M_\odot) < 6$ at $z \ge 10$. In our reference model, the latter population corresponds to heavy BH seeds formed by the direct collapse of super-massive stars in their earliest phases of mass growth \citep{trinca2022}. 

The first BH seeds are generally predicted to form in metal-poor or metal-free environments. Measuring the metallicity and stellar mass of the host galaxies of high redshift accreting BHs with JWST may provide additional important constraints on BH seeding models. 
We find that below $z < 10$ observable BHs are expected to be preferentially hosted by galaxies with metallicities ${\rm Log} (Z/Z_\odot) \ge -2$ and stellar masses $8 \leq {\rm Log} (M_*/M_\odot) < 10$, while at higher redshifts an increasing fraction of observable systems are expected to be hosted by galaxies with $-3 \leq {\rm Log} (Z/Z_\odot) < -2$ and lower stellar masses $6 \leq {\rm Log} (M_*/M_\odot) < 8$.

Our predictions are based on CAT (Cosmic Archaeology Tool), a semi-analytical galaxy/BH evolution model \citep{trinca2022}, which allows us to make ab initio predictions for BH seeds formation and growth, while being consistent with the general population of AGNs and galaxies observed at $4 \le z \le 7$. In the present study, we have based our predictions on CAT reference model, where the growth of light and heavy BH seeds can not exceed the Eddington rate. In this model, light BH seeds fail to grow due to inefficient gas accretion, and the observable BH population at $z > 5$ derives from grown heavy seeds and is characterized by accretion rates that fall in the range [0.1 - 1] $\dot{M}_{\rm Edd}$.  Observational constraints from the JWST survey will be precious to test this scenario.

Our predictions for the observability of the accreting BH population at $5 < z < 15$ with planned JWST surveys assume that these systems can be identified with properly designed colour selection criteria. In the past few years a number of these 
have been proposed,  based either on the continuum emission 
(see for example \citealt{Volonteri2017, natarajan2017, pacucci2017, valiante2018observability}), on emission-line selection \citep{Nakajima2022}, or on empirical templates \citep{Goulding2022}. 

Our study can potentially be expanded and improved in various directions. In particular,
CAT allows us to explore other model variants, such as the merger-driven model, where enhanced BH accretion episodes are triggered by galaxy mergers and can exceed the Eddington limit \citep{trinca2022}. In addition, we could base our predictions on a more sophisticated modelling of the BH Spectral Energy Distribution (SED) that depends on the black hole mass, accretion rate and on the metallicity of the host galaxy (see for instance the SED models proposed by \citealt{Volonteri2017, valiante2017, valiante2018observability}, \citealt{pezzulli2017a}, \citealt{Nakajima2022},  or \citealt{inayoshi2022}). The addition in CAT of dynamical channels forming intermediate mass BH seeds following \cite{sassano2021} could also provide a more complete census of massive BHs formed in high redshift galaxies. Finally, our estimates for the obscuration correction are based on an empirical model calibrated at $0.3 \le z \le 3$ \citep{merloni2014}. While nuclear BH obscuration may be more sensitive to the physical conditions prevailing in the nuclear region surrounding the BH rather than to the large-scale galactic environment, the physical properties of growing BH seeds may be different from their more massive and more luminous low-redshift descendants. Therefore, the effect of varying the obscuration as a function of BH properties and their environment may be an additional feature to investigate. 
In a future work we plan to provide a more extensive analysis along the lines described above and we will test colour selection criteria on the observable BH population predicted by our model.

\section*{Acknowledgements}
We acknowledge support from the Amaldi Research
Center funded by the MIUR program “Dipartimento di Eccellenza ”(CUP:B81I18001170001) and from the INFN TEONGRAV specific initiative. 
R.M. acknowledges support from the ERC Advanced Grant 695671, “QUENCH” and from the Science and Technology Facilities Council (STFC).
R.M. also acknowledges funding from a research professorship from the Royal Society.

\section*{Data Availability}
The simulated data underlying this article will be shared on reasonable request to the corresponding author.

\bibliographystyle{mnras}
\bibliography{Bibliography}

\bsp	
\label{lastpage}
\end{document}